\documentclass[aps,twocolumn,showpacs,amsmath,amssymb]{revtex4}
\usepackage{dcolumn}
\usepackage{bm}
\usepackage{color}
\usepackage{latexsym}
\usepackage{amssymb}
\usepackage{graphicx}
\usepackage{ulem}

\begin{document}
\title{Current instability and single-mode THz generation in ungated two-dimensional electron gas}

\author{M.V.Cheremisin, G.G.Samsonidze}

\affiliation{A.F.Ioffe Physical-Technical Institute,
St.Petersburg, Russia}

\begin{abstract}
We analyze the current instability of the steady state with a
direct current for an ungated two-dimensional(2D) electron layer
for arbitrary current and the carrier scattering strength. We
demonstrate the possibility of single-mode generator operating in
terahertz frequency range.
\end{abstract}

\maketitle

Recently, the interest in the plasma wave generation mechanism [1]
in an ultrashort-channel field effect transistor (FET) has
attracted much attention. The high-density 2D electron gas can be
described by equations similar to the hydrodynamic equations for
shallow water, resulting in the plasma wave generation [1]. Under
the asymmetric boundary conditions with a fixed voltage at the
source and fixed current at the drain lead to an instability of
the steady state with a dc current. The plasma wave instability in
short-channel FETs is very important for applications such as
high-power sources and both the non-resonant[2]-[4] and
resonant[5,6] detectors in the terahertz (THz) frequency range(see
also [7]. Recently, the current instability and resulting plasma
wave generation similar to that for field effect transistor shown
to occur in an ungated 2D electron layer for asymmetrical boundary
conditions [8].

In the present letter, we analyze the plasma wave instability for
arbitrary current and the carrier scattering strength. We
demonstrate the feasibility of single-mode generator operating in
terahertz frequency range.

We will use the hydrodynamic model of weakly damped($\omega \tau
\gg 1 $, where $\omega$ is the plasma frequency, $\tau$ is the
relevant collision time), incompressible charged-electron fluid
placed in a rigid, neutralizing positive background. Moreover, we
assume that the mean free path associated with electron-electron
collisions is less than that related to scattering by phonons and
impurities and than the device length. The behavior of the 2D
fluid is described [8] by the Euler and the continuity
one-dimensional equations:

\begin{eqnarray}
\frac{\partial v}{\partial t}+v\frac{\partial v}{\partial
x}=\frac{e}{m}\frac{\partial \phi}{\partial x}+\frac{v}{\tau}=0,
\label{Euler} \\
\frac{\partial n}{\partial t}+\frac{\partial(vn)}{\partial x}=0.
\label{continuity}
\end{eqnarray}
Here, $n$ is the 2D electron density, $v$ is the average electron
flux velocity, $m$ is the effective mass, $-e$ is the electronic
charge and $\phi$ is the potential, therefore the electric field
is $E=-\partial \phi/\partial x$.

Previous studies [1] correspond to dissipationless electrons when
$\tau \rightarrow \infty$. At small currents the steady electron
flow with a constant $v_{0},n_{0}$ and $\phi_{0}=0$ shown to be
unstable against small perturbations when the following boundary
conditions are fulfilled:
\begin{equation}
n(0,t)=n_{0},\qquad n(l,t)v(l,t)=j/We.
\label{boundary conditions}
\end{equation}
Here $n(0,t)$ is the fixed 2DEG density at the device
source($x=0$), $j$ is the fixed drain ($x=l$) current, and $W$ is
the 2D layer width. In the steady state, $n=n_{0}$ and
$v=v_{0}=j/(en_{0}W)$. These boundary conditions can be realized
[8] by grounding the source by large capacitance presenting a
short at plasma frequencies and by attaching the drain to the
power supply via an inductance that presents an open circuit at
the plasma frequencies.

We now search the solution of Eqs.(1) and (2) assuming finite
carrier dissipation. In this case the steady-state corresponds to
current carrying electron fluid with the constant density $n_{0}$(
because of local quasi- neutrality of 2D fluid ) and the constant
average velocity $v_{0}=\mu E=j/eWn_{0}$, where $\mu=e\tau/m$ is
the carrier mobility. Thus, the steady-state potential $\phi_{0}$
is linear downstream the electron flow. Let us now search the
evolution of small perturbations $v_{1}, n_{1},\phi_{1} \sim
\exp(-i\omega t+ikx)$ superimposed the steady state. Linearizing
Eqs.(1) and (2) with respect to $n_{1},v_{1}$ we find:
\begin{eqnarray}
(\omega+i/\tau-kv_{0})v_{1}=-\frac{e}{m}k\phi_{1}, \nonumber \\
(\omega-kv_{0})n_{1}=kn_{0}v_{1}. \label{linearizing}
\end{eqnarray}
Neglecting finite-size effects, we explore the relation
$\phi_{1}=-\frac{2\pi e n_{1}}{\left |k \right | \epsilon}$ known
for infinite ungated 2D electron layer. Here, $\epsilon$ is the
background dielectric constant. Therefore, the dispersion equation
yields:
\begin{equation}
(\omega+i/\tau-kv_{0})(\omega-kv_{0})=2a\left |k \right |,
\label{dispersion}
\end{equation}
where $a=\frac{\pi n_{0}e^{2}}{\epsilon m}$. Introducing the
dimensionless frequency
$\omega^{*}=\frac{v_{0}}{a}\left(\omega+\frac{i}{2 \tau}\right)$
we derive the dispersion relation for plasma wave propagating
upstream $k_{+}$ and downstream $k_{-}$ the current flow as
follows:
\begin{equation}
k_{\pm}=\pm \frac{1 \pm \omega^{*}-\sqrt{1 \pm
2\omega^{*}-\frac{v_{0}^{2}}{4\tau^{2}a^{2}} }}{v_{0}^{2}/a}.
\label{wave vectors}
\end{equation}
At small currents and zero-dissipation $1/\tau=0$ Eq.(6) reproduce
the dispersion relation $k_{\pm}=\pm \frac{\omega^{2}}{2a}\left (1
\mp \frac{\omega v_{0}}{a}\right )$ found in Ref.[8]. Searching
the solution of Eq.(4) in the form $n_{1}=A
\exp(ik_{+}x)+B\exp(ik_{-}x)$, and, then use the boundary
conditions(see Eq.(3)) for zero ac potential at the source
$n_{1}=0$ and zero ac current at the drain
$n_{0}v_{1}+v_{0}n_{1}=0$ we obtain
\begin{equation}
\frac{k_{1}}{k_{2}}=\exp \left[i({k_{+}-k_{-}})l \right],
\label{instability condition}
\end{equation}
Equation (6) and (7) allow one to determine both the real and
imaginary parts of the the complex frequency
$\omega=\omega'+i\omega''$. A positive imaginary part $\omega''
>0$ corresponds to instability. We now assume that for weakly damped
2D electrons the real part of the frequency is greater than both
the imaginary part and the dissipation term $1/\tau$. Retaining
the first order corrections in current $j \sim v_{0}$ in Eq.(6)
and, then, separating the real and imaginary parts of the
frequency we obtain:
\begin{eqnarray}
\Omega'_{n}=\frac{1}{2V}\left[1-\left(\frac{(2-V^{2}(2n-1))^{2}-2}{2}\right)^{2} \right]^{1/2}, \nonumber \\
\Omega''_{n}=\frac{V}{\pi}\frac{\sqrt{1-4(\Omega'_{n}V)^{2}}}{\sqrt{1-2\Omega'_{n}V}-\sqrt{1-2\Omega'_{n}V}}\ln
\left|R_{s}R_{d} \right |-\gamma, \nonumber \\
R_{s}=-1,
R_{d}=\frac{1-\Omega'_{n}V-\sqrt{1-2\Omega'_{n}V}}{1+\Omega'_{n}V-\sqrt{1-2\Omega'_{n}V}},
\label{frequency}
\end{eqnarray}
where we introduced the dimensionless frequency $\Omega=\omega
\sqrt{\frac{l}{\pi a}}$, carrier velocity $V=v_{0}
\sqrt{\frac{\pi}{a l}}$ and the instability damping strength
$\gamma=\frac{1}{2\tau}\sqrt{\frac{l}{\pi a}}$. Then, $R_{s},
R_{d}$ are the density amplitude reflection coefficients from the
source and drain boundaries, so that the product $R_{s}R_{d}$
denotes the instability gain factor. Note that for small currents
$V \rightarrow 0$ Eq.(8) describes the discreet frequency spectrum
$\omega'_{n}=\sqrt{\pi a(2n-1)/l}$ and mode-independent
instability increment $\omega''=v_{0}/l-1/2\tau$ reported in
Ref.[8].

\begin{figure}[tbp]\vspace*{0.5cm}
\includegraphics[scale = 0.7]{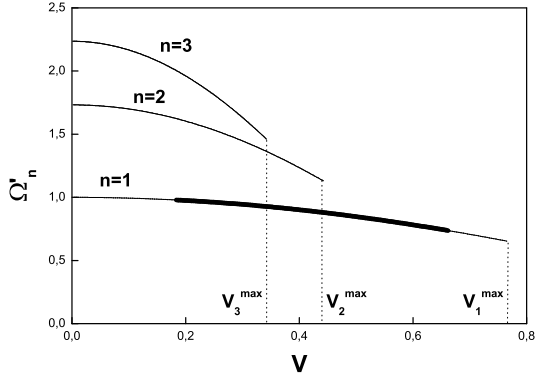}
\caption[]{\label{Fig1} Dimensionless frequency $\Omega'_{n}$ vs.
dimensionless carrier velocity $V$ for dissipationless electrons
$\gamma=0$ for $n=1,2,3$ modes. Vertical dotted lines denote the
cutoff $V^{max}_{n}$ of the instability increment(see Fig.2). Bold
line depicts the frequency of the single-mode generator at
$\gamma_{min}=0.055$} \vspace*{-0.5cm}
\end{figure}

The most intriguing result of the present paper concerns the
current dependence of $\Omega'_{n},\Omega''_{n}$. In order to
analyze the essential features of the plasma wave instability, we
first consider zero-dissipation electrons when $\gamma=0$. In
Fig.1,2 we plot the current dependence of the frequency and the
instability increment for the proper mode $n=1$ and the high-order
harmonics $n=2,3$. Each mode is characterized by its own increment
which is positive within a certain range of currents, i.e. when
$0<V<V^{max}_{n}=\sqrt{\frac{2-\sqrt{2}}{2n-1}}$. In Fig.1 we
represent the $n=1,2,3$ mode frequency within the respective range
of currents. Note that the higher the mode index $n$ the narrower
the n-th mode instability range. Actually, at $V=V^{max}_{n}$ the
n-th mode increment disappears because the related group velocity
of the upstream plasma wave $d\omega/dk_{-}$ vanishes( see Eq.6)).
Its worthwhile to mention that at $V>V^{max}_{2}=0.44$ the only
first mode remains unstable, hence the device may operate as a
single-mode generator.

\begin{figure}[tbp]\vspace*{0.5cm}
\includegraphics[scale = 0.7]{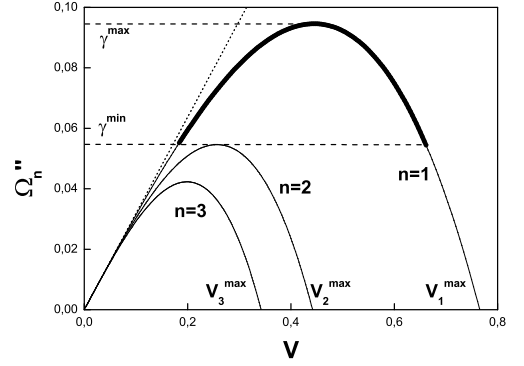}
\caption[]{\label{Fig2} Instability increment $\Omega''_{n},
n=1,2,3$ vs. dimensionless carrier velocity $V$ for
zero-dissipation electrons $\gamma=0$. The bold line depicts the
positive part of the instability increment for single-mode
generator when $\gamma=\gamma_{min}=0.055$. Dotted line represents
the zero-current asymptote $\omega''=v_{0}/l$ found in
Ref.\cite{Dyakonov05}} \vspace*{-0.5cm}
\end{figure}

In presence of finite dissipation the all modes are damped
identically( i.e. $\gamma=const$). Thus, one can find the
instability threshold for n-th mode as $\Omega''_{n}=0$. With the
help of Fig.2 we find the condition when the higher than the first
mode are suppressed:
\begin{eqnarray}
0.055=\gamma^{min}<\gamma<\gamma^{max}=0.095,\nonumber \\
0.183<V<0.661 \label{limits}
\end{eqnarray}
The instability is totally suppressed when $\gamma>\gamma^{max}$.
The frequency range of the single-mode generator for
$\gamma=\gamma^{min}$ is represented by the bold line in Fig.1.
Using Eq.(9) we now estimate the device parameters required for
observation of the single-mode plasma wave generation. For
$n=10^{12}$cm$^{-2}$ InGaAs based($m$=0.042$m_{e}$) 2D layer
length $l=1\mu$m we obtain the required range of 2D carrier
mobility as $3.2 \times 10^{4}$ cm$^{2}$/Vs$<\mu<5.5 \times
10^{4}$ cm$^{2}$/Vs. When $\gamma=\gamma^{min}$, the single-mode
generator output frequency and required carrier velocity fall in
the range $5.0$THz$<\omega'<6.8$THz and $4 \times 10^{5}<v_{0}<1.4
\times 10^{6}$cm/s respectively.

In conclusion we found the plasma wave instability for arbitrary
current and the carrier scattering strength. We demonstrate the
possibility of single-mode generator operating in terahertz
frequency range.


\begin{thebibliography}{100}
\bibitem{Dyakonov93}M.Dyakonov and M.Shur, Phys.Rev.Lett. {\bf 71}, 2465
(1993).
\bibitem{Dyakonov96}M.Dyakonov and M.Shur, IEEE Trans. Electron Devices
{\bf 43}, 380 (1996).
\bibitem{Lü}J.-Q. Lü, M.S.Shur, J.L.Hesler, L.Sun, and R.Weikle, IEEE Electron Device Lett.
{\bf 19}, 373 (1998).
\bibitem{Knap02a}W.Knap, V.Kachorovskii, Y.Deng, S.Rumyantsev, J.-Q.Lü, R.Gaska, M.S.Shur,
G.Simin, X.Hu, M.Asif Khan, C.A.Saylor, and L.C.Brunel, J. Appl.
Phys. {\bf 91}, 9346 (2002).
\bibitem{Knap02b}W.Knap, Y.Deng, S.Rumyantsev, J.-Q.Lü, M.S.Shur, C.A.Saylor, and L.C.Brunel,
Appl. Phys. Lett., {\bf 80}, 3433 (2002).
\bibitem{Knap02c}W.Knap, Y.Deng, S.Rumyantsev, and M.S.Shur, Appl. Phys. Lett., {\bf 80}, 4637 (2002)
\bibitem{Peralta}X.G.Peralta, S.J.Allen, M.C.Wanke, N.E.Narff, J.A.Simmons, M.P.Lilly, J.L.Reno,
P.J.Burke, and J.P.Eisenstein, Appl. Phys. Lett., {\bf 81}, 1627
(2002)
\bibitem{Dyakonov05}M.Dyakonov and M.Shur, Appl.Rhys.Lett. {\bf 87}, 111501 (2005).
\end{thebibliography}
\end{document}